\newcommand{\CLASSINPUTtoptextmargin}{19mm}%
\newcommand{\CLASSINPUTbottomtextmargin}{43mm}%
\newcommand{\CLASSINPUTinnersidemargin}{12.9mm}%
\newcommand{\CLASSINPUToutersidemargin}{12.9mm}%
\documentclass[conference,10pt,a4paper]{IEEEtran}
\usepackage{amsmath}
\usepackage{times}
\usepackage{graphicx}
\usepackage{multirow}
\usepackage[none]{hyphenat}
\usepackage{float}
\usepackage{subfig}
\usepackage{booktabs}

\usepackage{t1enc}
\usepackage{times}
\usepackage[table]{xcolor}
\usepackage{array}
\usepackage{eso-pic}

\makeatletter

\def\@maketitle{\newpage
\bgroup\par\addvspace{0.5\baselineskip}\centering%
\ifCLASSOPTIONtechnote
   {\bfseries\large\@IEEEcompsoconly{\sffamily}\@title\par}\vskip 1.3em{\lineskip .5em\@IEEEcompsoconly{\sffamily}\@author
   \@IEEEspecialpapernotice\par{\@IEEEcompsoconly{\vskip 1.5em\relax
   \@IEEEtitleabstractindextextbox{\@IEEEtitleabstractindextext}\par
   \hfill\@IEEEcompsocdiamondline\hfill\hbox{}\par}}}\relax
\else
   \vskip0.2em{\EuMWtitlesize\ifCLASSOPTIONtransmag\bfseries\LARGE\fi\@IEEEcompsoconly{\sffamily}\@IEEEcompsocconfonly{\normalfont\normalsize\vskip 2\@IEEEnormalsizeunitybaselineskip
   \bfseries\Large}\@title\par}\vskip1.0em\par
   \ifCLASSOPTIONconference%
      {\@IEEEspecialpapernotice\mbox{}\vskip\@IEEEauthorblockconfadjspace%
       \mbox{}\hfill\begin{@IEEEauthorhalign}\@author\end{@IEEEauthorhalign}\hfill\mbox{}\par}\relax
   \else
      \ifCLASSOPTIONpeerreviewca
         {\@IEEEcompsoconly{\sffamily}\@IEEEspecialpapernotice\mbox{}\vskip\@IEEEauthorblockconfadjspace%
          \mbox{}\hfill\begin{@IEEEauthorhalign}\@author\end{@IEEEauthorhalign}\hfill\mbox{}\par
          {\@IEEEcompsoconly{\vskip 1.5em\relax
           \@IEEEtitleabstractindextextbox{\@IEEEtitleabstractindextext}\par\hfill
           \@IEEEcompsocdiamondline\hfill\hbox{}\par}}}\relax
      \else
         \ifCLASSOPTIONtransmag
           {\@IEEEspecialpapernotice\mbox{}\vskip\@IEEEauthorblockconfadjspace%
            \mbox{}\hfill\begin{@IEEEauthorhalign}\@author\end{@IEEEauthorhalign}\hfill\mbox{}\par
           {\vspace{0.5\baselineskip}\relax\@IEEEtitleabstractindextextbox{\@IEEEtitleabstractindextext}\vspace{-1\baselineskip}\par}}\relax
         \else
           {\lineskip.5em\@IEEEcompsoconly{\sffamily}\sublargesize\@author\@IEEEspecialpapernotice\par
           {\@IEEEcompsoconly{\vskip 1.5em\relax
            \@IEEEtitleabstractindextextbox{\@IEEEtitleabstractindextext}\par\hfill
            \@IEEEcompsocdiamondline\hfill\hbox{}\par}}}\relax
         \fi
      \fi
   \fi
\fi\par\addvspace{0.0\baselineskip}\egroup}

\def\EuMWtitlesize{\@setfontsize{\EuMWtitlesize}{24}{24pt}}
\def\EuMWauthorsize{\@setfontsize{\EuMWauthorsize}{11}{11pt}}
\def\EuMWaffilsize{\@setfontsize{\EuMWaffilsize}{10}{10pt}}
\def\EuMWcaptionsize{\@setfontsize{\EuMWcaptionsize}{9}{10pt}}
\def\EuMWbibsize{\@setfontsize{\EuMWbibsize}{8}{10pt}}

\def\@IEEEauthorblockNstyle{\EuMWauthorsize\@IEEEcompsocnotconfonly{\sffamily}\@IEEEcompsocconfonly{\large}}
\def\@IEEEauthorblockAstyle{\EuMWaffilsize\@IEEEcompsocnotconfonly{\sffamily}\@IEEEcompsocconfonly{\itshape}\@IEEEcompsocconfonly{\large}}
\def\@IEEEauthordefaulttextstyle{\EuMWauthorsize\@IEEEcompsocnotconfonly{\sffamily}\sublargesize}

\def\thebibliography#1{\section*{\refname}%
    \addcontentsline{toc}{section}{\refname}%
    \EuMWbibsize\@IEEEcompsocconfonly{\small}\vskip 0.3\baselineskip plus 0.1\baselineskip minus 0.1\baselineskip
    \list{\@biblabel{\@arabic\c@enumiv}}%
    {\settowidth\labelwidth{\@biblabel{#1}}%
    \leftmargin\labelwidth
    \advance\leftmargin\labelsep\relax
    \itemsep \IEEEbibitemsep\relax
    \usecounter{enumiv}%
    \let\p@enumiv\@empty
    \renewcommand\theenumiv{\@arabic\c@enumiv}}%
    \let\@IEEElatexbibitem\bibitem%
    \def\bibitem{\@IEEEbibitemprefix\@IEEElatexbibitem}%
\def\newblock{\hskip .11em plus .33em minus .07em}%
\ifCLASSOPTIONtechnote\sloppy\clubpenalty4000\widowpenalty4000\interlinepenalty100%
\else\sloppy\clubpenalty4000\widowpenalty4000\interlinepenalty500\fi%
    \sfcode`\.=1000\relax}

\long\def\@makecaption#1#2{%
\ifx\@captype\@IEEEtablestring%
\par\@IEEEtabletopskipstrut
\else
\@IEEEfigurecaptionsepspace
\fi
\setbox\@tempboxa\hbox{\normalfont\footnotesize {#1.}\nobreakspace\nobreakspace #2}%
\ifdim \wd\@tempboxa >\hsize%
\setbox\@tempboxa\hbox{\normalfont\footnotesize {#1.}\nobreakspace\nobreakspace}%
\parbox[t]{\hsize}{\normalfont\footnotesize\noindent\unhbox\@tempboxa#2}%
\else
\ifCLASSOPTIONconference \hbox to\hsize{\normalfont\footnotesize\hfil\box\@tempboxa\hfil}%
\else \hbox to\hsize{\normalfont\footnotesize\box\@tempboxa\hfil}%
\fi\fi
\ifx\@captype\@IEEEtablestring%
\@IEEEtablecaptionsepspace
\else
\fi}

\newlength\tablecaptiontotableskip
\newlength\figuretocaptionskip
\def\@IEEEfigurecaptionsepspace{\vskip\figuretocaptionskip\relax}%
\def\@IEEEtablecaptionsepspace{\vskip\tablecaptiontotableskip\relax}%

\def\abstract{\normalfont%
\@IEEEabskeysecsize\bfseries\textit{\abstractname}\,\bfseries\textit{---}\,%
\@IEEEgobbleleadPARNLSP}%

\def\IEEEkeywords{\normalfont%
\@IEEEabskeysecsize\bfseries\textit{\IEEEkeywordsname}\,\bfseries\textit{---}\,%
\@IEEEgobbleleadPARNLSP}%
\def\endIEEEkeywords{\relax\vspace{0.67ex}%
\par\if@twocolumn\else\endquotation\fi%
\normalsize\normalfont}%

\def\@IEEEauthorblockNtopspace{0ex}
\def\@IEEEauthorblockAtopspace{1mm}
\def\tablename{Table}
\def\thetable{\arabic{table}}
\def\IEEEkeywordsname{Keywords}
\def\subsubsection{\@startsection{subsubsection}{3}{\z@}{1.5ex plus 1.5ex minus 0.5ex}%
{0.7ex plus .5ex minus 0ex}{\normalfont\normalsize\itshape}}%
\newlength{\CPheadmatchindent}%
\def\@seccntformat#1{\hbox to\CPheadmatchindent{\csname the#1dis\endcsname}\hskip 0.1em \relax}
\IEEEilabelindentA \parindent
\IEEEilabelindent \IEEEilabelindentA
\IEEEelabelindent \parindent
\IEEEdlabelindent \parindent
\IEEElabelindent \parindent
\makeatother

\DeclareUnicodeCharacter{24C7}{\textsuperscript{\textregistered}}
\begin{document}
\raggedbottom
%
%
%
\title{An Ultra-Compact Differential \textit{V}-Band Power Amplifier Using EDMOS Transistors With 18.1~dBm P\textsubscript{1dB} and 21\% PAE in 22nm FD-SOI CMOS}
%
%
\author{%
\IEEEauthorblockN{%
Han Zhou\textsuperscript{\protect\#\$}, Torgil Kjellberg\textsuperscript{\protect\&}, Haojie Chang\textsuperscript{\protect\#\$}, Christian Fager\textsuperscript{\protect\#}
}
\IEEEauthorblockA{%
\textsuperscript{\protect\#}Department of Microtechnology and Nanoscience,
Chalmers University of Technology, Sweden\\
\textsuperscript{\protect\$}Faculty of Information Technology and Communication Sciences,
Tampere University, Finland \\
\textsuperscript{\protect\&}Sivers Semiconductors AB, Sweden\\
han.zhou@tuni.fi
}
}
%
%
\AddToShipoutPictureFG*{%
  \AtPageUpperLeft{%
    \hspace{0.63in}%
    \raisebox{-0.32in}{%
      \parbox{\dimexpr\paperwidth-1.26in\relax}{%
        \centering\footnotesize\itshape
        This is the author-accepted version of a paper accepted for presentation at the
        2026 Asia-Pacific Microwave Conference (APMC 2026).
        \copyright~2026 IEEE. The final published version will be available in IEEE Xplore.
      }%
    }%
  }%
}
\maketitle
%
%
\begin{abstract}
This paper presents a compact, fully differential, two-stage millimeter-wave (mm-wave) cascode power amplifier (PA) designed and implemented in a 22nm FD-SOI CMOS process (22FDX+). The PA employs the newly introduced extended-drain MOS (EDMOS) device in 22FDX+, together with a carefully engineered device core and transformer baluns. At 50~GHz, the prototype achieves 18.8~dBm saturated output power (P\textsubscript{SAT}), 18.1~dBm 1-dB compression output power (P\textsubscript{1dB}), and 21\% power-added efficiency (PAE) at P\textsubscript{1dB}. To the best of our knowledge, this work achieves the highest reported power density of 2.6~W/mm\textsuperscript{2} among single-way, two-stage CMOS cascode PAs. 
\end{abstract}

%

\section{Introduction}
Driven by the increasing demand for high data throughput, wireless communication systems are moving toward higher mm-wave frequencies. Among them, the V-/E-band frequency range is particularly attractive due to its wide available bandwidth, low latency, and lightly licensed operation, making it well suited for high-speed wireless applications.

The power amplifier (PA) is a critical building block that strongly influences the performance of a wireless link. Its energy efficiency directly determines the overall system power consumption~\cite{intro}. With the increasing adoption of multi-antenna systems, such as multiple-input multiple-output (MIMO) and phased-array architectures, the output power requirement for each individual PA is relaxed, making CMOS PAs increasingly attractive. In such systems, power density ($\mathrm{W/mm^2}$) becomes a vital figure of merit in addition to absolute output power.

Device stacking and power combining are commonly used to increase output power. However, device stacking can degrade PA efficiency due to phase mismatches among stacked devices and additional routing losses. Power combining also introduces extra loss from passive combiners~\cite{BBDPAAI}. Therefore, a careful balance between moderate device stacking, such as a cascode configuration, and low-loss power-combining/routing is required to achieve a favorable performance tradeoff. The extended-drain MOS (EDMOS) device in the 22FDX+ CMOS process provides an attractive solution by enabling higher supply-voltage operation and, consequently, higher output power without efficiency degradation~\cite{EDMOS}. Although EDMOS transistors have been demonstrated in the 6G FR3 band~\cite{EDMOS6GFR3} and the 5G FR2 band~\cite{EDMOS5GFR2}, their implementation at $V$-band and higher millimeter-wave frequencies remains largely unexplored.

\begin{figure} [t!]
    \centering     
    \includegraphics[width=0.9\linewidth]{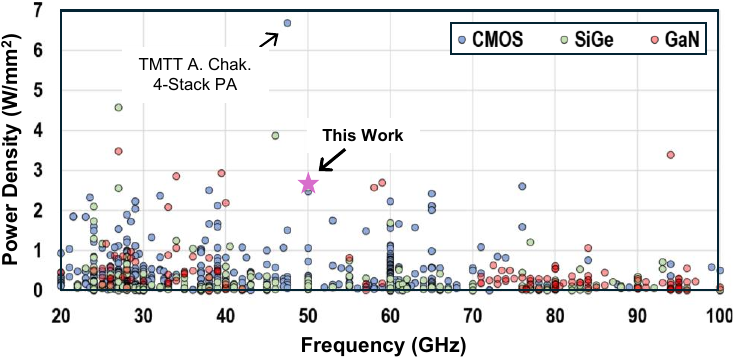}
    \caption{Comparison of power density, defined as $P_{\mathrm{sat}}$/core area, for PAs implemented in CMOS, SiGe, and GaN technologies across 20 to 100~GHz~[1].}
    \label{fig.1}
\end{figure}
\begin{figure} [t!]
    \centering        \includegraphics[width=0.62\linewidth]{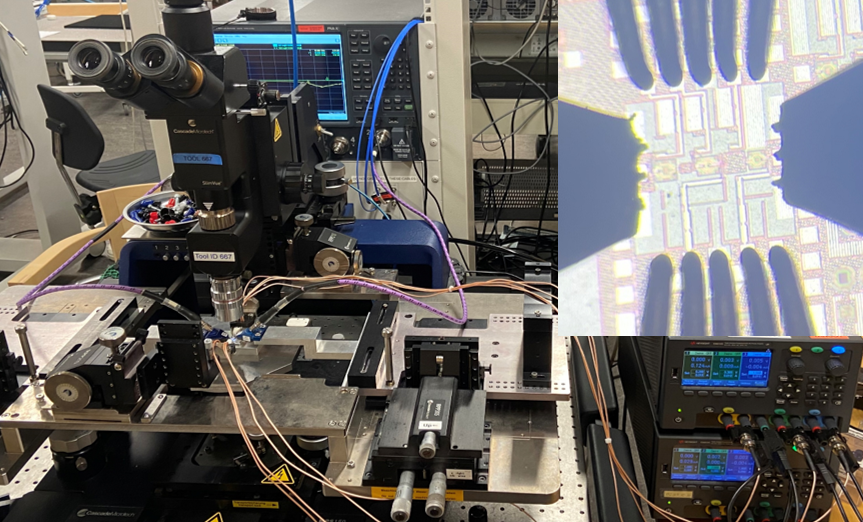}
    \caption{Probe measurement setup and chip micrograph of the fabricated PA.}
    \label{fig.2}
\end{figure}

In this paper, we analyze and design a fully differential, two-stage cascode PA that employs EDMOS transistors and a carefully engineered device core to enhance output power. Fig.~1 compares the power density of PAs implemented in CMOS, SiGe, and GaN technologies across 20 to 100~GHz~\cite{PA_Survey}. The fabricated prototype, with the measurement setup and chip photograph shown in Fig.~\ref{fig.2}, achieves a power density of $2.6~\mathrm{W/mm^2}$, which is among the highest reported values for CMOS cascode PA configurations.
\begin{figure} [t!]
    \centering     
    \includegraphics[width=0.95\linewidth]{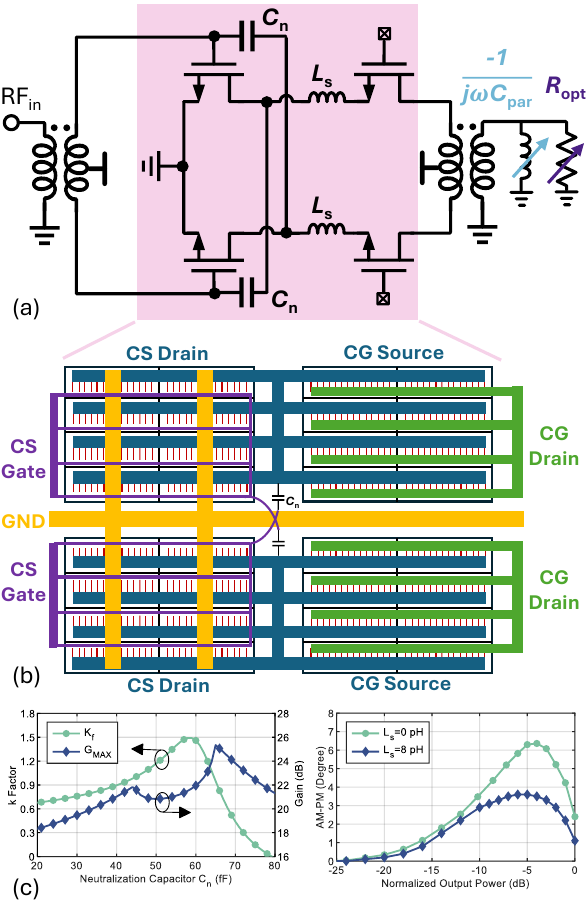}
    \caption{(a) Fully differential cascode PA simulation framework used to determine the neutralization capacitor ($C_n$), compensating inductor ($L_s$), and load-pull parameters ($R_{\mathrm{opt}}$ and $C_{\mathrm{par}}$). (b) Layout arrangement of the cascode PA transistor core for enhanced power performance. (c) Stability factor and maximum gain versus $C_n$, along with the PA AM–PM performance for different values of $L_s$.}
    \label{fig.3}
\end{figure}

\section{Analysis and Circuit Design}
As illustrated in Fig.~\ref{fig.3}(a), the PA design starts from a simplified schematic-level simulation setup. Taking the cascode stage as an example, a fully differential block is constructed using ideal baluns. This setup enables an initial estimation of the optimal load-pull parameters, including $R_{\mathrm{opt}}$ and $C_{\mathrm{par}}$, together with the transistor core size, neutralization capacitor $C_n$, and compensating inductor $L_s$.

\subsection{Device Core Selection and Design}
With the simulation setup shown in Fig.~\ref{fig.3}(a), the delivered output power, gain, PAE, and AM--PM performance are evaluated across frequency. Based on this initial analysis, the best tradeoff between output power and gain is obtained using unit EDMOS transistors with a gate finger width of $1~\mu\mathrm{m}$, a gate length of $24~\mathrm{nm}$, 16 gate fingers, and 2 vertical gate fingers. The unit EDMOS transistor is then carefully laid out. As demonstrated in Fig.~\ref{fig.3}(b), 16 unit devices are combined to form the common-source devices, while another 16 unit devices are combined to form the common-gate devices in the differential cascode stage. After completing the initial device-core routing, electromagnetic (EM) simulation of the core is performed. Subsequently, the neutralization capacitor and compensating inductor are swept at the design frequency, as shown in Fig.~\ref{fig.3}(c). Based on the sweep results, $C_n=58~\mathrm{fF}$ is selected to optimize gain and stability, while $L_s=8~\mathrm{nH}$ is chosen to improve AM--PM performance. The same design procedure is applied to the common-source driver stage, where 8 unit EDMOS devices are combined and a neutralization capacitor of $31~\mathrm{fF}$ is employed.
\begin{figure} [t!]
    \centering     
    \includegraphics[width=0.92\columnwidth]{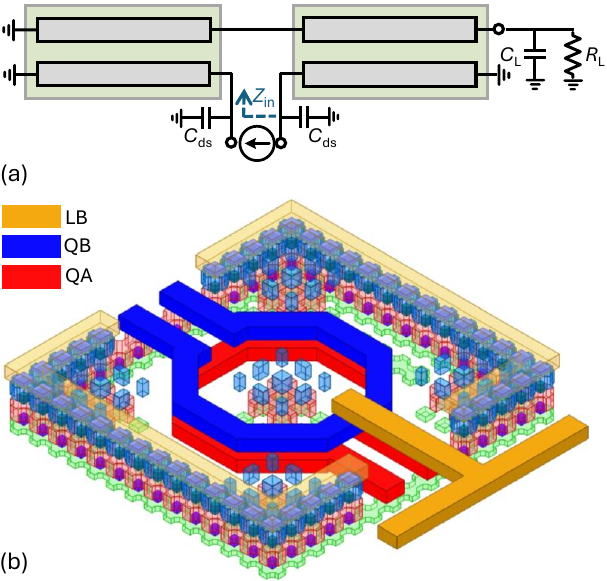}
    \caption{(a) Coupled-line balun/transformer synthesized using ideal current sources based on the obtained load-pull data. (b) Final layout of the output balun/transformer for the cascode PA stage.}
    \label{fig.4}
\end{figure}

\begin{figure*} [t!]
    \centering     
    \includegraphics[width=\textwidth]{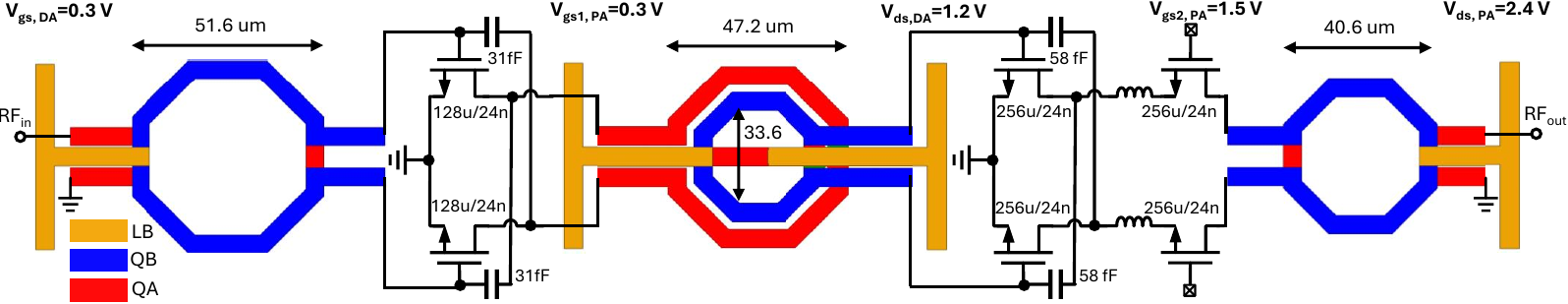}
    \caption{Schematic of the two-stage fully differential cascode PA implemented using EDMOS transistors in CMOS 22FDX+ technology.}
    \label{fig.5}
\end{figure*}

\subsection{Transformer and Balun Design}
After extracting the load-pull parameters of the cascode power stage and common-source driver stage, we design the output coupled-line balun following the method in~\cite{HW1, NoLMPA, 6GPP_Han}. As illustrated in Fig.~\ref{fig.4}(a), the coupled-line balun incorporates the transistor parasitic capacitance and pad capacitance into the overall matching network and presents the desired optimal impedance to the current source at the design frequency. Therefore, the extracted $R_{\mathrm{opt}}$ and $C_{\mathrm{par}}$ are used as simulation targets. In this design, they are $50~\Omega$ and $90~\mathrm{fF}$, respectively. Fig.~\ref{fig.4}(b) shows a 3-D overview of the final coupled-line balun. To ensure EM simulation accuracy, manual dummy fill is added inside the balun, while global dummy fill is excluded.

The input and inter-stage transformer baluns follow a more straightforward design procedure~\cite{TF1,RFDAC1}. First, we model the transformers analytically. For the input transformer, the small-signal input impedance of the driver stage is matched to $50~\Omega$. For the inter-stage transformer, we ensure that the output 1-dB compression point of the driver stage, after accounting for transformer loss and mismatch loss, remains higher than the input 1-dB compression point of the cascode power stage. The transformers are then EM-simulated to confirm good agreement with the analytical results.

\begin{figure} [t!]
    \centering     
    \includegraphics[width=0.6\columnwidth]{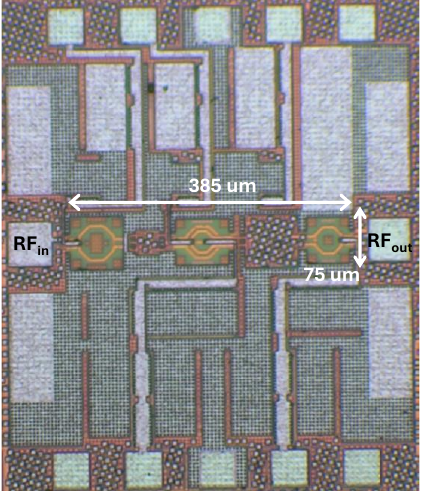}
    \caption{Die photograph of the fabricated chip.}
    \label{fig.6}
\end{figure}

\section{Circuit Implementation}
Fig.~\ref{fig.5} shows the schematic of the designed PA, including the dimensions of the input, inter-stage, and output transformer baluns. The driver stage uses a common-source transistor with a total gate width of $128~\mu\mathrm{m}$ and operates from a $1.2~\mathrm{V}$ supply. The power stage employs a cascode topology, where both the common-source and common-gate transistors have a total gate width of $256~\mu\mathrm{m}$ and operate from a $2.4~\mathrm{V}$ supply. 

Fig.~\ref{fig.6} shows the microphotograph of the fabricated PA implemented in GlobalFoundries 22nm FDX+ technology. The prototype occupies a compact core area of $0.028~\mathrm{mm^2}$.

\begin{table*}[t]
\centering
\caption{Comparison With State-of-the-Art High-Efficiency PAs}
\label{tab:pa_comparison}
\begin{tabular}{
|>{\columncolor{gray!15}\bfseries\boldmath}c
|>{\columncolor{blue!8}}c
|c|c|c|c|c|}
\hline
\rowcolor{gray!25}
 & \cellcolor{blue!15}\textbf{This Work} & \textbf{RFIC 2024~\cite{TB1}} & \textbf{MWTL 2024~\cite{TB2}} & \textbf{MWTL 2025~\cite{EDMOS6GFR3}} & \textbf{EuMIC 2023~\cite{TB4}} & \textbf{TMTT 2014~\cite{TB5}} \\ \hline

Technology &
\begin{tabular}[c]{@{}c@{}}22-nm FD-SOI\\ CMOS\end{tabular} &
\begin{tabular}[c]{@{}c@{}}22-nm FD-SOI\\ CMOS\end{tabular} &
\begin{tabular}[c]{@{}c@{}}22-nm FD-SOI\\ CMOS\end{tabular} &
\begin{tabular}[c]{@{}c@{}}22-nm FD-SOI\\ CMOS\end{tabular} &
\begin{tabular}[c]{@{}c@{}}150-nm GaN\\ on SiC\end{tabular} &
\begin{tabular}[c]{@{}c@{}}45-nm SOI\\ CMOS\end{tabular} \\ \hline

Architecture &
\begin{tabular}[c]{@{}c@{}}2-stage, cascode,\\ 1-way\end{tabular} &
\begin{tabular}[c]{@{}c@{}}1-stage, CS,\\ 1-way\end{tabular} &
\begin{tabular}[c]{@{}c@{}}1-stage, cascode,\\ 1-way\end{tabular} &
\begin{tabular}[c]{@{}c@{}}2-stage, cascode,\\ 1-way\end{tabular} &
\begin{tabular}[c]{@{}c@{}}2-stage, CS,\\ 2-way\end{tabular} &
\begin{tabular}[c]{@{}c@{}}2-stage, 4-stack,\\ 1-way\end{tabular} \\ \hline

Supply Voltage (V) &
1.2 / 2.4 & 1.1 & 1.6 & 0.8 / 2.4 & 25 / 25 & 2.4 / 4.8 \\ \hline

Frequency (GHz) &
50 & 60 & 76 & 12 & 62 & 47 \\ \hline

Gain (dB) &
16.2 & 15.1 & 12.3 & 25.2 & 9.3 & 24.9 \\ \hline

$P_{\mathrm{sat}}$ (dBm) &
18.8 & 14.3 & 14.6 & 23.5 & 31.3 & 20.1 \\ \hline

$P_{\mathrm{1dB}}$ (dBm) &
18.1 & 10.6 & 10.1 & 22.8 & 28.5$^{*}$ & 16.7$^{*}$ \\ \hline

PAE$_{\mathrm{1dB}}$ (\%) &
21 & 20$^{*}$ & 15$^{*}$ & 38 & 9$^{*}$ & 11$^{*}$ \\ \hline

Core Area ($\mathrm{mm^2}$) &
0.028 & 0.020 & 0.018 & 0.152 & 4.5$^{\dagger}$ & 0.016 \\ \hline

\begin{tabular}[c]{@{}c@{}}
\bfseries\boldmath Power Density\\
\bfseries\boldmath Based on $P_{\mathrm{sat}}$ / $P_{\mathrm{1dB}}$\\
\bfseries\boldmath ($\mathrm{W/mm^2}$)
\end{tabular} &
2.6 / 2.2 & 1.3 / 0.6 & 1.6 / 0.6 & 1.5 / 1.3 & 0.3 / 0.16 & 6.4 / 2.9 \\ \hline

\end{tabular}

\vspace{1mm}
\footnotesize{$^{*}$Estimated from reported plots or available data. $^{\dagger}$Including the pads.}
\end{table*}

\section{Measurement Results}
The fabricated prototype is mounted on a copper plate for characterization. DC and RF probes are used on a probe station to bias and characterize the PA. Small-signal measurements are performed using a Keysight PNA-X, while continuous-wave (CW) large-signal measurements are conducted using an external signal generator and power meter.

\begin{figure} [t!]
    \centering     
    \includegraphics[width=0.85\columnwidth]{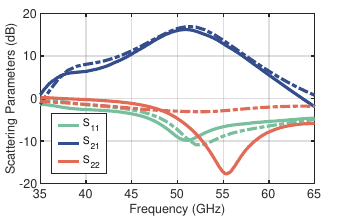}
    \caption{Measured (solid line) and simulated (dashed line) small-signal results of the fabricated PA prototype.}
    \label{fig.7}
\end{figure}

\begin{figure} [t!]
    \centering     
    \includegraphics[width=0.85\columnwidth]{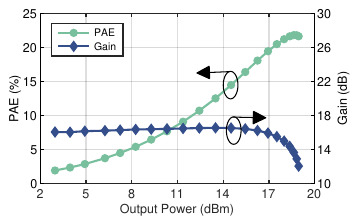}
    \caption{Measured PAE and gain versus output power at $50$~GHz.}
    \label{fig.8}
\end{figure}

\begin{figure} [t!]
    \centering     
    \includegraphics[width=0.85\columnwidth]{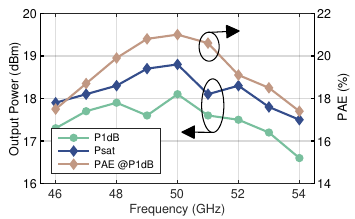}
    \caption{Measured saturated output power, 1-dB compression point ($P_{\mathrm{1dB}}$), and PAE at $P_{\mathrm{1dB}}$ of the prototype versus frequency.}
    \label{fig.9}
\end{figure}

\subsection{Small-Siganl S-Parameter Measurements}
The measured S-parameters of the fabricated prototype show good agreement with the simulation results, as shown in Fig.~\ref{fig.7}. The measured 3-dB small-signal bandwidth spans from $47.1$ to $53.8~\mathrm{GHz}$. At the center frequency, the PA achieves a small-signal gain of $16.2~\mathrm{dB}$. When biased closer to class-AB operation, the measured small-signal gain can increase to $21~\mathrm{dB}$, at the expense of slight degradation in efficiency and AM--AM performance. A small discrepancy between the measured and simulated gain is observed near the edges of the design band.

\subsection{Large-Signal CW Measurements}
Fig.~\ref{fig.8} presents the CW measurement results at $50~\mathrm{GHz}$. The PA achieves $18.8~\mathrm{dBm}$ saturated output power ($P_{\mathrm{sat}}$), $18.1~\mathrm{dBm}$ 1-dB compression output power ($P_{\mathrm{1dB}}$), and $21\%$ PAE at $P_{\mathrm{1dB}}$. Fig.~\ref{fig.9} shows the measured PA performance across frequency. From $46$ to $54~\mathrm{GHz}$, $P_{\mathrm{1dB}}$ remains above $16.5~\mathrm{dBm}$, and the PAE at $P_{\mathrm{1dB}}$ remains above $17.6\%$.

\section{Conclusion}

This paper presents a compact, two-stage $V$-band cascode PA in GlobalFoundries 22nm FDX+ technology. The design employs EDMOS transistors, a carefully engineered device core, and transformer-balun co-design with device and pad parasitics to achieve high output power and efficiency within a compact core area. The fabricated prototype occupies only $0.028~\mathrm{mm^2}$ and achieves a small-signal gain of $16.2~\mathrm{dB}$ at $50~\mathrm{GHz}$, with a 3-dB bandwidth from $47.1$ to $53.8~\mathrm{GHz}$. We further validate the PA through CW large-signal measurements, where it achieves a saturated output power of $18.8~\mathrm{dBm}$, a $P_{\mathrm{1dB}}$ of $18.1~\mathrm{dBm}$, and a PAE of $21\%$ at $P_{\mathrm{1dB}}$. Across $46$ to $54~\mathrm{GHz}$, the PA maintains a $P_{\mathrm{1dB}}$ above $16.5~\mathrm{dBm}$ and a PAE at $P_{\mathrm{1dB}}$ above $17.6\%$. With a power density of $2.6~\mathrm{W/mm^2}$ based on $P_{\mathrm{sat}}$, this work achieves one of the highest reported power-density values among CMOS cascode PA configurations.

\section*{Acknowledgment}
This research was supported in part by Swedish Innovation Agency (VINNOVA), Sivers Semiconductors, and Chalmers University of Technology under Grant 2022-00863, and in part by VINNOVA Grant 2024-02531, MULTIRACS, through the Eureka CELTIC Framework.


\bibliographystyle{IEEEtran}

\bibliography{IEEEabrv,mybibfile}

\end{document}